\documentclass[aps, prl, twocolumn]{revtex4-1}
\usepackage{graphicx}
\usepackage{subfigure}
\usepackage{amsmath}
\usepackage[bookmarks, hyperindex, colorlinks=true,
            linkcolor=red,
            urlcolor=blue, citecolor=blue]{hyperref}

\usepackage{bm}
\usepackage[usenames,dvipsnames,svgnames,table]{xcolor}

\renewcommand{\vec}[1]{\bm{#1}}
\newcommand{\subfigimg}[3][,]{%
  \setbox1=\hbox{\includegraphics[#1]{#3}}
  \leavevmode\rlap{\usebox1}
  \rlap{\hspace*{47.5pt}\vspace*{45pt}\raisebox{\dimexpr\ht1-2\baselineskip}{#2}}
  \phantom{\usebox1}
}

\begin{document}
\title{The importance of $\sigma$ bonding electrons for the accurate description of electron correlation in graphene}
\author{Huihuo Zheng, Yu Gan, Peter Abbamonte, and Lucas K. Wagner}
\email{lkwagner@illinois.edu}
\affiliation{
Department of Physics, University of Illinois at Urbana-Champaign,  Urbana, Illinois 61801-3080, USA}
\date{\today}

\begin{abstract}
Electron correlation in graphene is unique because of the interplay of the Dirac cone dispersion of $\pi$ electrons with long range Coulomb interaction. 
The random phase approximation predicts no metallic screening at long distance and low energy because of the zero density of states at Fermi level, so one might expect that graphene should be a poorly screened system. 
However, empirically graphene is a weakly interacting semimetal, which leads to the question of how electron correlations take place in graphene at different length scales.
We address this question by computing the equal time and dynamic structure factor $S(\vec q)$ and $S(\vec q, \omega)$ of freestanding graphene using {\it ab-initio} fixed-node diffusion Monte Carlo and the random phase approximation. 
We find that the $\sigma$ electrons contribute strongly to $S(\vec q,\omega)$ for relevant experimental values of $\omega$ even at distances up to around 80 \AA.
These findings illustrate how the emergent physics from underlying Coulomb interactions results in the observed weakly correlated semimetal.
\end{abstract}
\maketitle

Graphene has drawn much attention in the last decade because of its unusual electronic and structural properties and its potential applications in electronics~\cite{Wallace1947, Novoselov2004,NovoselovNature2005, Katsnelson2006, Geim2007, Novoselov2007, neto2009, Castro2009}. 
Although many electronic properties of graphene can be successfully described in a noninteracting electron picture~\cite{Castro2009}, electron-electron interactions do play a central role in a wide range of electronic phenomena that have been observed in experiments~\cite{Kotov2012}. For example, the reshaping of the Dirac cone was first predicted~\cite{Gonzalez1996, Steven2001, DasSarma2007} and later observed experimentally~\cite{Reshape2011}.
The fractional quantum hall effect has been observed under high magnetic field~\cite{Bolotin2009}. Collective plasmon and plasmaron excitation have also been observed~\cite{Bostwick2010,Bassani1967,Marinopoulos2004,Eberlein2008}. 

The interplay of the effective Dirac cone with Coulomb interactions makes the correlation effects in graphene unique. The random phase approximation (RPA) predicts no screening in graphene at long distance and static limit because the density of states is zero at the Fermi level. This is in contrast to normal metals, where charge carriers and impurities are highly screened by the Fermi sea through a formation of virtual electron-hole pairs according to RPA~\cite{Pines1997}. 
Therefore, it is an interesting question how electronic response takes place in graphene and how to describe it accurately.

In recent years, it has become possible to obtain very high resolution inelastic X-ray (IXS) experiments on graphite~\cite{Reed2010,Gan2015}, which were then modified to obtain information about the graphene planes.
These experiments directly probe the dynamical structure factor $S(\vec q,\omega)$, which allows for a detailed look at the electron correlations.
The main purpose of these experiments was to investigate the role of screening at long wave lengths, and particularly whether the random phase approximation (RPA) obtains an accurate representation of the physics at long range.
While these studies obtained unprecedented details for the low-energy charge excitations, their interpretation is challenging because of limited experimental resolution and uncertainties about the reference for RPA; whether the $\sigma$ bonding electrons are included or not, and what the underlying theory is, have large effects on the result~\cite{Exciton2010}.
For small enough wave vector $q$ and small enough energy $\omega$, the effect of the $\sigma$ electrons should be small, but it is unclear whether the experiments have reached that regime.

In this manuscript, we address both the question of the suitability of RPA perturbation theory and the effect of $\sigma$ electrons by applying highly accurate first-principles diffusion quantum Monte Carlo (DMC) to a series of planar systems including graphene.
DMC is a non-perturbative method with minimal approximations \cite{QMC,Kolorenc2010,Spink2013} and explicit representation of the electron-electron interactions. 
It has been shown to be a highly accurate method on both molecular systems and solids~\cite{QMC, Qwalk, WagnerDipole2007, Kolorenc2010, Spink2013, Drummond2013,wagner_effect_2014, Zheng2015, Wagner2015}.
We compute the structure factor $S(\vec q)$ which gives information about the long-range density-density correlations in the material and compare it directly to the X-ray data, obtaining agreement within the experimental error bars.
We find that the bonding $\sigma$ electrons are surprisingly important even at the longest range accessible to experiment, and if the RPA is performed including the $\sigma$ electrons from a DFT reference, it is in good agreement with the experimental data, although the peak locations do depend on the reference as noted previously~\cite{Exciton2010}.

The structure factor $S(\vec q)$ is a measure of the equal-time charge-charge correlations of the system, defined as
\begin{eqnarray}\label{eq:sq}
S(\vec q) :=\frac{1}{N}\langle \rho_{-\vec q}\rho_{\vec q}\rangle, 
\end{eqnarray} 
where $N$ is the number of electrons in the system, and $\rho_{\vec q} = e^{i\vec q\cdot  \hat{ \vec r}}$ is the density operator in reciprocal space.  
$S(\vec q)$ is directly related to the Coulomb interactions of the system \cite{Chiesa2006},
\begin{eqnarray}
V = \frac{e^2}{4\pi^2}\int d\vec q \frac{S(\vec q)-1}{q^2},
\end{eqnarray}
where $V$ is the Coulomb energy per particle. 
$S(\vec q)$ is the integral of the dynamic structure factor $S(\vec q, \omega)$ over frequency domain:
\begin{eqnarray}\label{eq:sqsqw}
S(\vec q) &=& \int_{0}^{\infty} \frac{d\omega}{2\pi} S(\vec q, \omega)=\frac{\hbar \vec q^2}{2m}\frac{\int_0^{\infty} d\omega S(\vec q, \omega) }{\int_{0}^{\infty} d\omega \omega S(\vec q, \omega)}.
\end{eqnarray}
Here we have applied the f-sum rule for $S(\vec q, \omega)$ \cite{Pines1997}.
The dynamic structure factor describes the dielectric response of the system \cite{Pines1997}. 
It can be directly measured through inelastic X-ray scattering \cite{Reed2010}, and can also be computed using RPA \cite{Pines1997,  Mowbray2014}. 

The first-principles calculations were performed as follows. DFT calculations were first performed using the CRYSTAL package~\cite{CRYSTAL} with Perdew-Burke-Ernzerhof (PBE) exchange and correlation functional \cite{PBE}. The simulations were performed on a $16\times 16$ supercell including 512 atoms. 
Burkatzki-Filippi-Dolg (BFD) pseudopotentials \cite{BFD2007, BFD2008} were used to remove the core electrons. 
The result of the DFT calculations is a Slater determinant made of Kohn-Sham orbitals. 
The Slater determinant was then multiplied by a Jastrow correlation factor and optimized using variance optimization \cite{QMC}. 
DMC calculations were performed using the QWalk package \cite{Qwalk} to obtain $S(\vec q)$. 
RPA calculations were performed using the GPAW package~\cite{Mortensen2005,Yan2011} to obtain $S(\vec q, \omega)$.
The Hubbard model was solved by auxiliary-field quantum Monte Carlo method (AFQMC) using the QUEST package~\cite{Varney2009}. 
\begin{figure}[hbt]
\centering
\includegraphics[clip, width=0.85\linewidth]{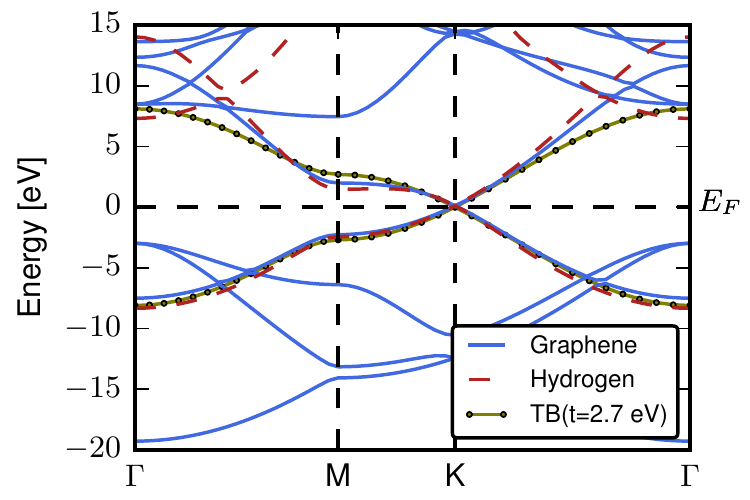}
\caption{Band structure of graphene, hydrogen and tight-binding model (with nearest-neighbor hopping $t=2.7$eV).}
\label{fig:band}
\end{figure}
\begin{table}[ht]
\caption{Systems/models investigated}\label{tab:sm}
\centering
\begin{tabular}{|c|c|c|}
\hline
system/model  & electrons & ~~~~~method ~~~~~\\
\hline
\hline
 Graphene (G): a=$2.46$ \AA$^{-1}$ & $\sigma$ \& $\pi$ & DMC, S-J\footnote{S-J: $S(\vec q)$ is evaluated on a Slater-Jastrow wavefunction through variational Monte Carlo using QWalk package. The Slater determinant is formed by the occupied $\pi$ and $\sigma$ orbitals. }, RPA\\
 $\pi$-only graphene (G$_{\pi}$)\footnote{G$_\pi$: $S(\vec q)$ is evaluated on a Slater-Jastrow wave function. The Slater determinant is formed by only the occupied $\pi$ orbitals but without including $\sigma$ orbitals. } & $\pi$ & S-J \\
 Hydrogen (H): a=$2.46$ \AA$^{-1}$ & s & DMC, RPA \\
 Tight-binding (TB): $t=2.7$ eV\footnote{TB: The value of $t$ has been chosen to match the DFT band dispersion near Dirac point. } & $\pi$ & RPA \\
 Hubbard: $U/t=1.6$\footnote{Hubbard: Hubbard model on a honeycomb lattice with onsite interation. } & $\pi$ & AFQMC \\
\hline
\end{tabular}
\end{table}

In order to disentangle different contributions to $S(\vec q, \omega)$ from $\pi$ and $\sigma$ electrons in graphene, we compared $S(\vec q)$ among the five systems listed in Table~\ref{tab:sm}.
All systems have similar low energy band structure  (see Fig.~\ref{fig:band}),  but differ in the presence or absence of $\sigma$ electrons, and in the interaction between electrons. 
The s orbital of the hydrogen lattice has almost the same dispersion as the $\pi$ orbital in graphene (see Fig.~\ref{fig:band}), which provides a way to understand the behavior of $\pi$ electrons in graphene in the absence of $\sigma$ electrons while still retaining a $1/r$ interaction. 
The graphene and hydrogen systems are studied using DMC and RPA. 
The tight-binding model ($t$=2.7eV) is studied using RPA with $1/r$ interactions. $S(\vec q)$ is obtained by integration of $S(\vec q, \omega)$ according to Eq.~\eqref{eq:sqsqw}.

\begin{figure}[hbt]
\centering
\includegraphics[clip, width=0.85\linewidth]{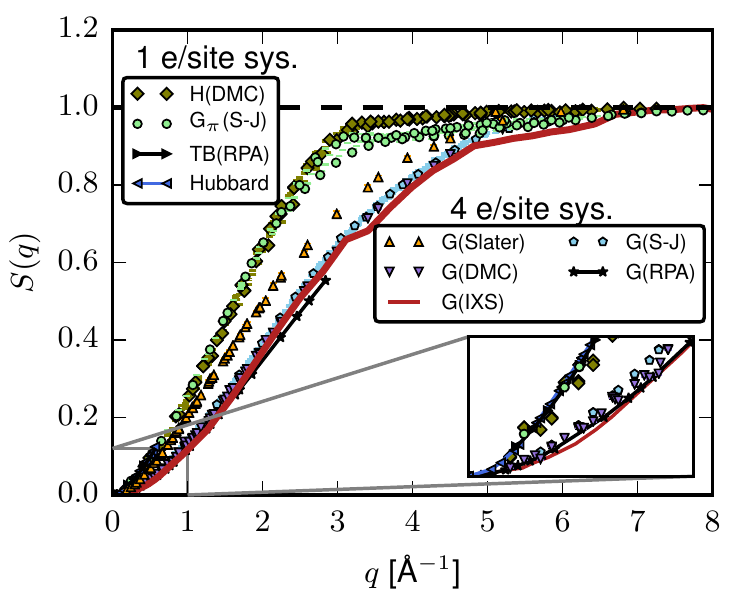}
\caption{$S(q)$ of different systems obtained through different methods. G: graphene; G$_{\pi}$: $\pi$ electrons only graphene; H: hydrogen; TB: $\pi$-band tight-binding model with $1/r$ interactions; Hubbard: the Hubbard model with $U/t = 1.6$. }
\label{fig:fit_curve}
\end{figure}

Let us first consider the $S(\vec q)$ results for {\it ab-initio} graphene, denoted by G in Fig.~\ref{fig:fit_curve}. 
For comparison, we have plotted $S(\vec q)$ of a non-interacting Slater determinant of Kohn-Sham orbitals [G(Slater)], and that of a Slater-Jastrow wavefunction [G(S-J)].
Both RPA and DMC results are very close to the experimental IXS results, but there is a significant difference between the correlated calculations and the Slater determinant, as expected.
A Slater-Jastrow wavefunction is indistinguishable from DMC results.
It thus appears that the experimental $S(\vec q)$ is well reproduced by any of these three correlation techniques (RPA, S-J and DMC) for {\it ab-initio} wave functions.
Quantitatively, this change of $S(\vec q)$ from the Slater determinant reflects a reduction of the Coulomb energy by $1.31(5)$ eV per electron.

Now consider the models H, TB, Hubbard, and $G_{\pi}$ in Fig.~\ref{fig:fit_curve}, which only contain one electron per site, in contrast to the four electrons per site of {\it ab-initio} graphene.
Each of these models has a computed $S(\vec q)$ quite close to the others.
Therefore, regardless of the computational method and interaction, if one electron per site is considered, $S({\vec q})$ is about the same, while for four electrons per site, $S({\vec q})$ is about the same if a correlated method is used.
We can thus assess the importance of the lower energy electrons, the $\sigma$ electrons, at different wavelengths by comparing the one electron per site curves to the four electron per site curves.

At the smallest values of $q$ available to both the computational and experimental techniques, the  $1$e/site $S({\vec q})$ differs from the $4$e/site $S({\vec q})$. 
If it were the case that the $\sigma$ electrons did not contribute to the long-range density density fluctuations, we would expect those $S({\vec q})$ values to coincide for $q$ small enough.
We thus conclude that the $\sigma$ electrons contribute to the density density fluctuations even for $q\sim 0.1-0.2$\AA$^{-1}$ and that the equal time $S({\vec q})$ is accurately described by RPA, Slater-Jastrow, and diffusion Monte Carlo.

\begin{figure*}[tbh]
\centering
  \begin{tabular}{@{}p{0.9\linewidth}@{\quad}p{\linewidth}@{}}
    \subfigimg[clip]{(a)}{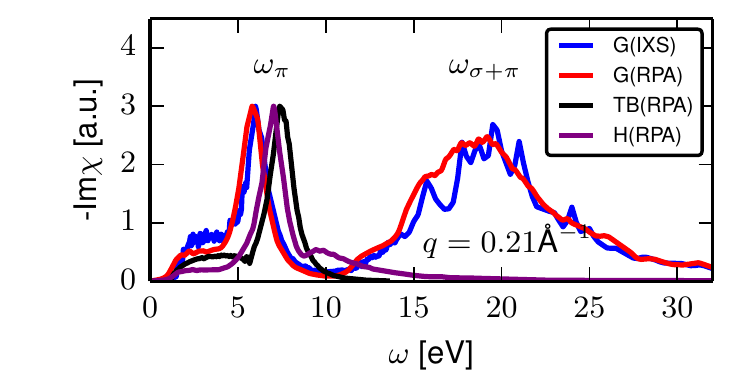}
    \subfigimg[clip]{(b)}{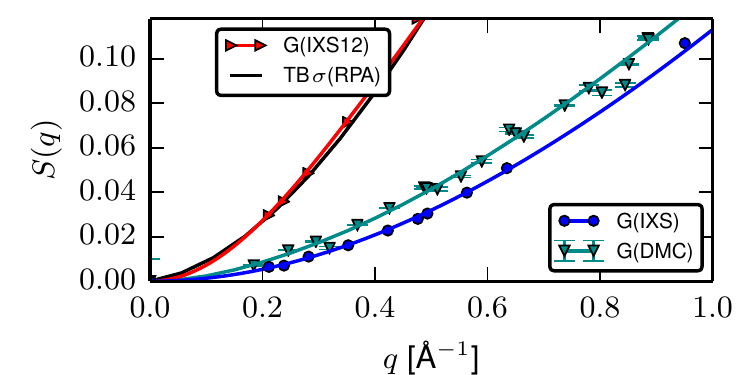}\\
    \subfigimg[clip]{(c)}{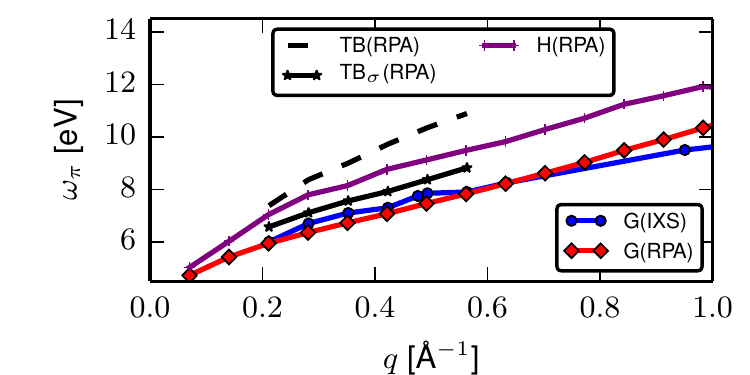}
    \subfigimg[clip]{(d)}{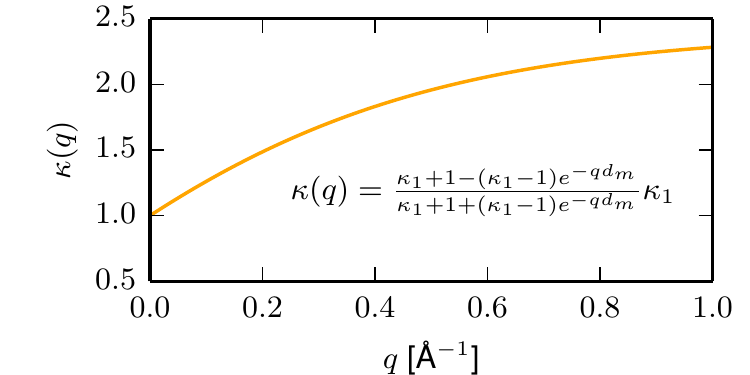}
      \end{tabular}
\caption{Effect of $\sigma$ electrons at long range. (a) imaginary part of the response function $\chi(\vec q, \omega)$ of graphene (IXS and RPA), tight-binding model with $1/r$ interaction, and hydrogen lattice at different momentum transfer. -Im$\chi$ has included local field corrections, and has been scaled in order to compare with experiment. The different curves have been scaled for comparison. (b) $S(q)$ from integration of $S(q, \omega)$ with different energy cutoff $\omega_{c}$. (c) dispersion of $\pi\to\pi^{*}$ plasmon resonance peaks. (d) effective screening function from $\sigma$ electrons. }
\label{fig:eels}
\end{figure*}

Now let us move to the dynamic response of graphene, $S(\vec q,\omega)$. 
Fig.~\ref{fig:eels}(a) shows the computed imaginary part of the response function $\chi(\vec q, \omega)$ of graphene in comparison with the X-ray experiment. 
There are mainly two resonance peaks in the spectrum~\cite{Bassani1967, Ahuja1997, Eberlein2008}: (1) $\pi \to \pi^*$ interband transition ($\pi$  plasmons); (2) $\sigma\to\pi^*$ and $\pi \to \sigma^*$ interband transition ($\sigma+\pi$ plasmons), denoted as $\omega_{\pi}$ and $\omega_{\sigma+\pi}$ respectively in Fig.~\ref{fig:eels}(a). 
RPA accurately reproduces the experimental IXS data, although the level of agreement for RPA may be partially fortuitous \cite{Exciton2010}. 
This is also why the $S(q)$ of G(RPA) matches G(DMC) and G(IXS) very well as is shown in Fig.~\ref{fig:fit_curve}, from which we know that there are no large peaks that are missing in RPA. 
It thus appears that the long-range response, at least to the limits of experimental resolution and potentially with small errors in the peak positions, is well-described by RPA calculations that include the $\sigma$ electrons.

The $\pi+\sigma$ plasmons are inherently missing in the TB model and hydrogen lattice. 
This is why $S(\vec q)$ in these two systems is larger than that in graphene as is shown in Fig.~\ref{fig:fit_curve}.
To see this, let us recall how we compute $S(\vec q)$ from the X-ray measured $S(\vec q, \omega)$ using Eq.~\eqref{eq:sq}.
The frequency cutoff in our experimental data is $2,000$ eV which is high enough to include relevant excitations of valence electrons. 
If we include  only $\pi$ plasmons but exclude $\sigma + \pi$ plasmons in the integration of Eq.~\eqref{eq:sq}, by setting energy cutoff to be $12$eV. 
The $S(q)$ matches the $S(q)$ of the TB model [see the G(IXS12) curve in Fig.~\ref{fig:eels}(d)].

The $\pi$ plasmon resonance peak $\omega_{\pi}$ of the TB model and hydrogen is $1\sim 3$ eV larger than that of graphene [Fig.~\ref{fig:eels}(a) and (d)]. 
This indicates strong screening effects from $\sigma$ electrons in graphene which are not present in TB model nor in hydrogen system. 
The interaction from $\sigma$ electron ``renormalizes" the $\pi$ plasmon resonance frequency. 

Let us investigate the ``renormalization" effect from $\sigma$ electrons by reconsidering the dynamic response of TB model. The TB(RPA) curves in Fig.~\ref{fig:eels} are computed by assuming that the tight-binding model is put in vacuum and that the $\pi$ electrons interact with each other through a bare Coulomb interaction ($1/r$). Suppose now it is in an environment of $\sigma$ electrons, within the RPA framework, we would have to include a background dielectric function $\kappa_{\sigma}(\vec q)$ in the response function $\chi(\vec q, \omega)$ \cite{Yuan2011, Gan2015}, 
\begin{eqnarray}\label{eq:chi_sigma}
\chi(\vec q, \omega) &=& \frac{\Pi_\text{TB}(\vec q, \omega)}{\kappa_{\sigma}(\vec q)-V(\vec q)\Pi_\text{TB}(\vec q, \omega)}.
\end{eqnarray}
$\Pi_\text{TB}(\vec q, \omega)$ is the original polarization function of TB model computed using the Lindhard function~\cite{Yuan2011}, and $V(\vec q)$ is the Fourier transformation of Coulomb interation. A good estimation of $\kappa_{\sigma}$ is \cite{Wehling2011, Yuan2011} [plotted in Fig.~\ref{fig:eels}(d)] , 
\begin{eqnarray}\label{eq:kappa}
\kappa_{\sigma}(\vec q) &=& \frac{\kappa_{1} + 1 - (\kappa_{1} - 1)e^{-qL}}{\kappa_{1} + 1 + (\kappa_{1} - 1)e^{-qL}}\kappa_{1}\,.
\end{eqnarray}
$\kappa_{1}\simeq 2.4$ is the dielectric constant of graphite, and $L=2.8$\AA ~is the effective thickness for a single layer graphene. 
The inclusion of $\kappa_\sigma$ indeed reduces $\omega_\pi$ by about $1\sim 2$ eV [see the TB$_{\sigma}$(RPA) curve in Fig.~\ref{fig:eels}(b)]. Thus, we have clearly demonstrated in  the RPA level that the screening from $\sigma$ electrons reduces the $\pi$ plasmon resonance frequency. 

\begin{figure}[tbh]
\centering
\includegraphics[clip, width=0.85\linewidth]{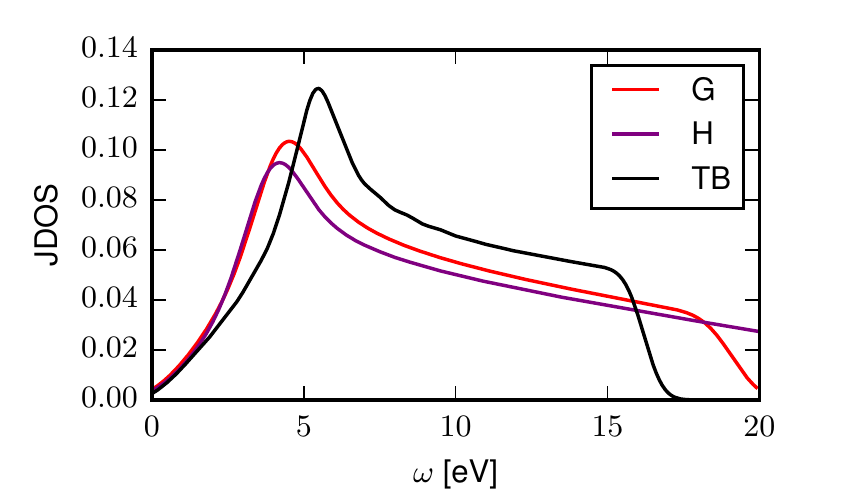}
\caption{Joint density of states (JDOS) of graphene (G), hydrogen (H), and tight-binding (TB) model. The JDOS of graphene and hydrogen are computed from PBE band structure.}\label{fig:jointdos}
\end{figure}

The remaining discrepancy between TB$_{\sigma}$(RPA) and G(IXS) [or G(RPA)] is partially due to the deviation of tight-binding band dispersion from \textit{ab-initio} graphene, since the $\pi$-plasmon frequency $\omega_\pi$ is directly related to the $\pi \to \pi^*$ interband transition energy.  The band structure deviation can be seen from the joint density of states, 
\begin{eqnarray}
jdos(\omega) = \frac{1}{N_k}\sum_{\vec k} \delta(\epsilon_{\pi^{*}}(\vec k) - \epsilon_{\pi}(\vec k) - \omega)\,,
\end{eqnarray}
where $\epsilon_{\pi}(\vec k)$ and $\epsilon_{\pi^{*}}(\vec k)$ are the eigenvalues of $\pi$ band and $\pi^{*}$ band respectively, and $N_k$ is the number of $k$ points sampled in the first Brillouin zone. The joint density of graphene, hydrogen and tight-binding model are shown in Fig.~\ref{fig:jointdos}. Because of the the van Hove singularity at the M point, there is a peak located near $\epsilon_{\pi^{*}}(M) - \epsilon_{\pi}(M)$. The $\pi\to\pi^{*}$ transition energy at M point is $5.4$ eV for the tight-binding model, $4.2$ eV for \textit{ab-initio} graphene, and $4.0$ eV for hydrogen (see Fig.~\ref{fig:band}). 
The peak shifts towards high energy from graphene to the tight-binding model,  which causes a change of $\omega_{\pi}$ by about $1$ eV. Therefore,  one needs to take into account of the band structure deviation when comparing results from tight-binding model with that of graphene. 

Now consider the screening effect from $\sigma$ electrons at different wavelengths. This can be seen from the difference of $\omega_\pi$ between graphene and hydrogen system regarding the fact that these two systems have very close band structure up to the energy scale of $\omega_\pi$ [see the joint density of states in Fig.~\ref{fig:jointdos}]. 
This difference of $\omega_\pi$ in the two systems decreases as $q \to 0$ [Fig.~\ref{fig:eels} (c)]. 
In the limit as $q \rightarrow 0$, the response from the $\sigma$ electrons goes to zero, as is also reflected in $\kappa_{\sigma}$: $\kappa_{\sigma}\to 1$ if $q \to 0$, and $\kappa_{\sigma} \to \kappa_{1}$ for large $q$ [see Fig.~\ref{fig:eels}(d)].  
However, even at the lowest momentum that the X-ray experiment has access to ($q=0.21$\AA$^{-1}$) \cite{Gan2015}, the screening effect is not small ($\kappa_{\sigma}(q) \simeq 1.5$). 
This causes a shift of $\omega_\pi$ by $0.8$ eV, which is comparable to estimations of excitonic effects \cite{Reed2010, Exciton2010, Gan2015}. 
If it is desired to isolate the $\pi$ electrons from the $\sigma$ electron screening, $|q| \ll 0.21$\AA$^{-1}$ must be accessed. For example, at $q = 0.07$ \AA $^{{-1}}$, the shift reduces to $0.25$ eV [see Fig.~\ref{fig:eels}(c)]. 

In conclusion, using the first-principles quantum Monte Carlo approach and the random phase approximation with DFT as the reference, we are able to describe the electron correlation in graphene accurately and reproduce the X-ray data very well for all $q$ available, provided that the $\sigma$ electrons are included in the calculations. 
The level of agreement for RPA may be fortuitous \cite{Exciton2010}, but it is clear that the $\sigma$ electrons are important for the interpretation of IXS data even at ranges up to around 80~\AA. 
Quantum Monte Carlo as a check on RPA and experiment confirms this fact.  
For very small values of $\omega$, the $\pi$-only model is accurate, but the experimental data does not reach those regimes. 

The $\sigma$ electrons affect the calculation in two important ways. 
First, at long wavelength, the $\sigma$ electrons respond through $\pi+\sigma$ plasmons which causes graphene to have a $S(q)$ different from tight-binding model and hydrogen. 
The screening from $\sigma$ electrons reduces the $\pi$ plasmons resonance frequency for about $1\sim 2$ eV comparable to other effects that will cause similar shift such as excitonic effects \cite{Exciton2010}. 
Second, with the presence of $\sigma$ electrons, the band structure of graphene deviates from a $\pi$-orbital tight-binding model, which further modifies the spectrum. These effects are observable even at $q=0.21$\AA $^{-1}$, the lowest momentum that current X-ray experiments can access. 

This study shows that unprecedented detail into electron correlations can be obtained both from the experimental and theoretical points of view, even for a system like graphene which has an unusual low-energy band structure. 
Without adjustable parameters, we have demonstrated the direct correspondence between density-density fluctuations measured by inelastic X-ray experiments and that calculated by theory. 
If the effects of all valence electrons in graphene are carefully taken into account, it appears possible to account for most of the correlations in graphene using standard techniques. 

The authors would like to thank Hitesh Changlani for helpful discussions in performing AFQMC calculations for Hubbard model, Andr\'e Schleife for useful discussions, Bruno Uchoa for inspiring the work, and Duncan J. Mowbray for providing RPA calculations details for graphene. 
This material is based upon work supported by the U.S. Department of Energy, Office of Science, Office of Advanced Scientific Computing Research, Scientific Discovery through Advanced Computing (SciDAC) program under Award Number FG02-12ER46875. Computational resources were provided by the DOE INCITE SuperMatSim/PhotoSuper programs and the Taub campus cluster at the University of Illinois at Urbana-Champaign.

\bibliography{screening}
\end{document}